\documentclass[12pt]{article}

\parskip=\medskipamount                         
\def\7#1#2{\mathop{\null#2}\limits^{#1}}        

\def\beee{\begin{equation}}
\def\eeee{\end{equation}}

\usepackage{graphicx}
\usepackage{setspace}
\usepackage{amssymb}
\usepackage[T1]{fontenc}
\newcounter{myfootertablecounter}
\newcommand\myfootnotemark{%
\addtocounter{footnote}{1}%
\footnotemark[\thefootnote]%
}%
\newcommand\myfootnotetext[1]{%
\addtocounter{myfootertablecounter}{1}
\footnotetext[\value{myfootertablecounter}]{#1} }
\newcommand\myfootnote[1]{%
\addtocounter{myfootertablecounter}{1} \footnote{#1} }%
\begin{document}
\begin{center}
 \textbf { \large Possible Detection of Causality
Violation in a Non-local
Scalar Model}\\
[5mm] Asrarul Haque\myfootnote{email address:ahaque@iitk.ac.in}\\
{ Department of Physics, I.I.T. Kanpur, Kanpur 208016(INDIA)}\\
Satish D. Joglekar\myfootnote{email address: sdj@iitk.ac.in}\\
{ Department of Physics, I.I.T. Kanpur, Kanpur 208016(INDIA)}\\
and\\
NISER, Bhubaneswar 751005 (INDIA)
\end{center}
\begin{abstract}
We consider the possibility that there may be causality violation
detectable at higher energies. We take a scalar nonlocal theory
containing a mass scale $\Lambda$ as a model example and make a
preliminary study of how the causality violation can be observed.
We show how to formulate an observable whose detection would
signal causality violation. We study the range of energies
(relative to $\Lambda$) and couplings to which the observable can
be used.
\end{abstract}
\section{{\textsc{\normalsize introduction}}}
Non-local quantum field theories (NLQFT) have been a subject of
wide research since 1950's. The main reason for the interest in
early days has been the hope that the non-local quantum field
theory can provide a solution to the puzzling aspects of
renormalization. The basic idea was that since the divergences in
a local quantum filed theory arise from product of fields at
identical space-time point, the divergences of the local quantum
field theory would be tamed if the interaction were non-local. In
particular, if the interaction scale was typically of the order of
$1/\Lambda$, then momenta in loop integrals (Euclidean) would be
damped when $|p^{2}|>>\Lambda^{2}$. The early work on NLQFT,
starting from that by Pais and Uhlenbeck \cite{PU} and especially
that of Efimov and coworkers, has been summarized in \cite{NA}.
NLQFT's also have found application towards description of
extended particles which incorporates the symmetries of the theory
in some (non-local) form \cite{M90}. The non-commutative fields
theories, currently being studied \cite{NC}, are a special variant
of a NLQFT, as is evident especially in its QFT representation
using the star product. In this work, we shall focus our attention
on the type of NLQFT's formulated by Kleppe and Woodard
\cite{KW92}. One of the reasons we normally insist on a\textbf{
}\emph{local}\textbf{ }quantum field theory is because it has
micro-causality, and this generally ensures causality of the
theory. One of the consequences, therefore, that would be
suspected of non-locality would be a causality violation at the
level of the S-matrix. Indeed, since at a given moment, the
interaction is spread over a finite region in space, thus covering
simultaneously space-like separated points, we expect the
interaction to induce non-causality. In view of the fact that we
have not observed large-scale causality violation, it becomes
important to distinguish between theories exhibiting\textbf{
}\emph{classical}\textbf{ }violations of causality versus\textbf{
}\emph{quantum}\textbf{ }violations of causality. As argued in
\cite{J06}, a violation of causality at the classical level can
have a larger effective range and strength, compared to the
quantum violations of causality which are suppressed by
~$g^{2}/16\pi^{2}$ per loop. We do not know of large scale
causality violations, and as such, it is desirable that the
non-local theory has no classical violation of causality. One way
known to ensure that there is no classical level of causality
violation is to require that the S-matrix of the NLQFT at the tree
level coincides with that of the local theory
($\Lambda\rightarrow\infty$) as is arranged in the formulation of
\cite{KW92}. We shall work in the context of the NLQFT's as
formulated by Kleppe and Woodard \cite{KW92}. This form of
non-local QFT was evolved out of earlier work of Moffat
\cite{M90}, insights into structure of non-local field equations
by Eliezer and Woodard \cite{EW} and application to QED by Evens
et al \cite{E91}. This formulation has a distinct advantage over
earlier attempts in several ways:
\begin{enumerate}
\item There are no additional classical solutions to the non-local field
equations compared to the local ones. The nonlocal theory is truly
a deformation of the local theory and the meaning of quantization,
as a perturbation about the classical, is not altered. This
property is not shared by non-commutative field theories.
\item It has the same S-matrix at the tree level, and thus;
\item There is no classical violation of causality.
\item The theory, unlike a higher derivative theory, has no ghosts and is
unitary at a \emph{finite $\Lambda$}.
\item The theory can embody non-localized versions of local symmetries having
an equivalent set of consequences.
\end{enumerate}
There are many other reasons for taking interest in these NLQFT's.
We have found such a non-local formulation \emph{with a finite
$\Lambda$,} very useful in understanding the renormalization
program in the renormalizable field theories \cite{j01}. We have
shown that this formulation enables one to construct a
mathematically consistent framework in which the renormalization
program can be understood in a natural manner. The framework does
not require any violations of mathematical rigor usually
associated with the renormalization program. This framework,
moreover, made it possible to theoretically estimate the mass
scale $\Lambda$. The nonlocal formulations can also be understood
\cite{j01_2} as an effective field theory formulation of a
physical theory that is valid up to mass scale $\sim$$\Lambda$. In
such a case, the unknown physics at energy scales higher than
$\Lambda$ {[}such as a structure in terms of finer constituents,
additional particles, forces, supersymmetry etc ] can
\emph{effectively be represented} in a \emph{consistent} way (a
unitary, gauge-invariant, finite (or renormalizable) theory) by
the non-local theory. In other words, the nonlocal standard model
can serve as such an effective field theory \cite{j01_2} and will
afford a model-independent way of consistently reparametrizing the
effects beyond standard model. It can be looked upon in a number
of other ways. One could think the non-locality as representing a
form factor with a momentum cut-off $\Lambda$\cite{M90}. One could
also think of this theory as embodying a granularity of space-time
of the
scale $1/\Lambda$ or as an intrinsic mass scale $\Lambda$ \cite{KW92,js,j01}.\\
 A possible \char`\"{}limitation\char`\"{} of the theory is that the
theory necessarily has quantum violations of causality
\cite{KW92,CO92}; though it can be interpreted as a prediction of
the theory. In another work, A. Jain and one of us explored the
question with the help of the simple calculations for the simplest
field theory: the nonlocal version of the $\lambda\phi^{4}$ theory
\cite{JJ}. While, in this scalar field model, the causality
violation is related to the nonlocality of interaction put in by
hand, so to speak, in practice such a non-locality of interaction
could arise from many possible sources. It could arise from a
fundamental length, $1/\Lambda$, present in nature. It could arise
from composite nature of elementary particles (This possibility
has recently been explored \cite{J08}). In this work, we wish to
formulate how the effect can be observed experimentally. In order
to study causality violation (CV) in the theory, it is first
necessary to formulate quantities that signal CV. We would like to
construct quantities that can be\textbf{ }\emph{measured}\textbf{
}experimentally. From this view-point{\footnote{there are, of course, results based on dispersion relation approach}}, it is appropriate to
construct quantities in terms of the S-operator. Bogoliubov and
Shirkov \cite{BS} have formulated a\textbf{
}\emph{necessary}\textbf{ }condition for causality to be preserved
in particle physics by the S-operator . This formulation is simple
and at the same time extremely general in that, it uses only (i)
the phenomenologically accessible S-operator together with (ii)
the most basic notion of causality in a relativistic formulation:
A cause at $x$ shall not affect physics at any point $y$ unless
$y$ is in the forward light-cone with respect to $x$. The
condition is formulated as,\begin{equation}
\begin{array}{ccc}
\frac{\delta}{\delta g(x)}\left(\frac{\delta S[g]}{\delta
g(y)}S^{\dagger}[g]\right)=0 & for & x<\sim
y\end{array}\label{eq:CC}\end{equation}
 where $x<\sim y$ means that either $x^{0}<y^{0}$ or $x$ and $y$
are space like separated. {[}In either case, there exists a frame
in which $x^{0}<y^{0}$]. Section~\ref{sec:2} gives a brief
qualitative understanding of this relation and how amplitudes
indicating causality violation are constructed using this
relation. In section \ref{sec:2}, we shall also summarize the essentials of construction of a non-local QFT given a local one. In this section, we shall give the results for the exclusive processes $\phi+\phi\rightarrow\phi+\phi$ in the one loop order from \cite{JJ}. In section 3, we make a comparison of the local contribution and the non-local CV effects and the latter could be significant for $s\leq\Lambda^{2}$ and when one analyzes angular distributions. In section 4, we shall construct a physical observable in terms of a differential cross-section $\frac{d\sigma}{d\Omega}$. This quantity involves some higher order terms and in section 6, we shall make an estimate of them and show that under certain conditions on coupling constant and energies they are indeed negligible and allow observation of the observable constructed in section 3.

While, what we have presented for simplicity, is a model calculation, similar attempt can be made for a more realistic process in the standard model. A work, along the same lines, but applicable to the realistic cases of experimentally observed exclusive processes $e^{+}e^{-}\rightarrow e^{+}e^{-}$,$ e^{+}e^{-}\rightarrow\mu^{+}\mu^{-}$and $e^{+}e^{-}\rightarrow\tau^{+}\tau^{-}$ is in progress.

\section{{\normalsize \textsc{preliminary} \label{sec:2}}}
In this section, we shall briefly discuss the construction of
non-local field theories and the Bogoliubov-Shirkov criterion of
causality. We shall further summarize results on causality
violation calculation in \cite{JJ,J06}.
\subsection{\textmd{\textsc{\normalsize non-local quantum field theory}}}
We shall present the construction of the NLQFT as presented in
\cite{KW92}. We start with the local action for a field theory, in
terms of a generic field $\phi$, as the sum of the quadratic and
the interaction part:
\[
S[\phi]=F[\phi]+I[\phi]\]
 and express the quadratic piece as \[
F[\phi]=\int d^{4}x\phi_{i}(x)\Im_{ij}\phi_{j}(x)\] We define the
regularized action in terms of the smeared field $\widehat{\phi}$,
defined in terms of\myfootnote{The choice of the smearing operator
is the only freedom in the above construction. For a set of
restrictions to be fulfilled by $\mathcal{E}$, see e.g.
\cite{CO92} } the kinetic energy operator $\Im_{ij}$ as,\[
\widehat{\phi}\equiv\mathcal{E}^{-1}\phi\,\,\,\,\,\,\,\,\,\,\,\mathcal{E}\equiv\exp[\Im/\Lambda^{2}]\]
 The nonlocally regularized action is constructed by first introducing
an auxiliary action $S[\phi,\psi]$. It is given by \[
S[\phi,\psi]=F[\hat{\phi}]-A[\psi]+I[\phi+\psi]\]
 where $\psi$ is called a {}``shadow field'' with an action\[
A[\psi]=\int d^{4}x\psi_{i}O_{ij}^{-1}\psi_{j};\;\;
O\equiv\frac{\mathcal{E}^{2}-1}{\Im}\]
 The action of the non-local theory is defined as $\hat{S}[\phi]=S[\phi,\psi]\Vert_{_{\psi=\psi[\phi]}}$
where $\psi[\phi]$ is the solution of the classical equation
$\frac{\delta S}{\delta\psi}=0$.\\ The vertices are unchanged but
every leg can connect either to a smeared propagator\[
\frac{i\mathcal{E}^{2}}{\Im+i\epsilon}=-i\int_{1}^{\infty}\frac{d\tau}{\Lambda^{2}}exp\{\frac{\Im\tau}{\Lambda^{2}}\}\]
 or to a shadow propagator {[}shown by a line crossed by a bar]
\[
\frac{i[1-\mathcal{E}^{2}]}{\Im+i\epsilon}=-iO=-i\int_{0}^{1}\frac{d\tau}{\Lambda^{2}}exp\{\frac{\Im\tau}{\Lambda^{2}}\}\]
 In the context of the $\lambda\phi^{4}$ theory, we have, \[
\Im=-\partial^{2}-m^{2}\quad I(\phi)=-\frac{g}{4}\phi^{4}\]
 We shall now make elaborative comments. The procedure constructs
an action having an infinite number of terms (each individually
local), and having arbitrary order derivatives of $\phi$. The net
result is to give convergence in the\textbf{
}\emph{Euclidean}\textbf{ }momentum space beyond a momentum scale
$\Lambda$ through a factor of the form
$\exp{(\frac{p^{2}-m^{2}}{\Lambda^{2}})}$ in propagators. The
construction is such that there is a one-to-one correspondence
between the solutions of the local and the non-local classical
field equations, (a difficult task indeed \cite{EW}). It can also
be made to preserve the local symmetries of the local action in a
non-localized form \cite{KW92}. The Feynman rules for the scalar
nonlocal theory are simple extensions of the local ones. In
momentum space, these read:
\begin{enumerate}
\item For the $\phi$-propagator (smeared propagator) denoted by a straight
line: \[ i\frac{
exp\left[\frac{p^{2}-m^{2}+i\epsilon}{\Lambda^{2}}\right]
}{p^{2}-m^{2}+i\epsilon}=\frac{-i}{\Lambda^{2}}\int_{1}^{\infty}d\tau\exp\left\{
\tau\left[\frac{p^{2}-m^{2}+i\epsilon}{\Lambda^{2}}\right]\right\}
\]
\item For the $\psi$-propagator denoted by a\textbf{ }\emph{barred}\textbf{
}line:\[ i\frac{
1-exp\left[\frac{p^{2}-m^{2}+i\epsilon}{\Lambda^{2}}\right]
}{p^{2}-m^{2}+i\epsilon}=\frac{-i}{\Lambda^{2}}\int_{0}^{1}d\tau\exp\left\{
\tau\left[\frac{p^{2}-m^{2}+i\epsilon}{\Lambda^{2}}\right]\right\}
\]
\item The 4-point vertex is as in the local theory, except that any of the
lines emerging from it can be of either type. (There is
accordingly a statistical factor).
\item In a Feynman diagram, the internal lines can be either shadow or smeared,
with the exception that no diagrams can have closed shadow loops.
\end{enumerate}
Lower bound has been put on the scale of non-locality
$\Lambda$\cite{js,AJ} from $g-2$ of muon and precision tests of
standard model. It has been argued that an upper bound on the
scale $\Lambda$ can be obtained from the requirement that
renormalization program is naturally understood in a nonlocal
field theory setting \cite{j01,j01_2}. Should particles of
standard model be composite, $\Lambda$ could naturally be related
to the compositeness scale \cite{J08}.
\subsection{\textmd{\textsc{\normalsize bogoliubov-shirkov causality criterion}}}
The causality condition that we have used to investigate causality
violation in NLQFT is the one discussed by Bogoliubov and Shirkov
\cite{BS}. They have shown that an S-matrix for a theory that
preserves causality must satisfy the condition of Eq.(\ref{eq:CC})
\begin{equation}
\frac{\delta}{{\delta g(x)}}\left({\frac{{\delta S(g)}}{{\delta
g(y)}}S^{\dagger}(g)}\right)=0~for~x<\sim
y\label{eq:CC}\end{equation} and it has been formulated treating
the coupling $g(x)$ as space-time dependent. A simple\textbf{
}\emph{qualitative}\textbf{ }understanding can be provided as in
\cite{J06}. The above relation is a series in $g(x)$ and leads
perturbatively to an infinite set of equations when expanded using
\begin{equation} S[g]=1+\sum_{n\geq1}\frac{1}{n!}\int
S_{n}(x_{1},...,x_{n})g(x_{1})...g(x_{n})dx_{1}...dx_{n}.\label{eq:CCexp}\end{equation}
 We consider the following expression \begin{eqnarray*}
H(y;g) & = & i\frac{{\delta S(g)}}{{\delta g(y)}}S^{\dagger}(g)\\
 & = & \sum\limits _{n\ge0}{\frac{1}{{n!}}}\int{H_{n}(y,x_{1},...x_{n})g(x_{1})...g(x_{n})dx_{1}...dx_{n}}\end{eqnarray*}
We shall write only a few of each of these coefficient functions
\begin{equation} H_{1}(x,y)\equiv
iS_{2}(x,y)+iS_{1}(x)S_{1}^{\dagger}(y)\label{causal1}\end{equation}
\begin{equation}
H_{2}(x,y,z)\equiv
iS_{3}(x,y,z)+iS_{1}(x)S_{2}^{\dagger}(y,z)+iS_{2}(x,y)S_{1}^{\dagger}(z)+iS_{2}(x,z)S_{1}^{\dagger}(y)\label{causal2}\end{equation}
(for a general expression for $H_{n}$, see \cite{JJ}). Then, the
causality condition (\ref{eq:CC}) reads,
\[
\frac{\delta}{{\delta g(x)}}H\left(y,g\right)=0\qquad for\; x<\sim
y\] which implies,\begin{equation} H_{1}\left(x,y\right)=0\qquad
y<\sim x\label{eq:causal1}\end{equation}
\begin{equation}
H_{2}\left(x,y,z\right)=0\qquad y<\sim
x\,\mbox{and/or}\,z<\sim x \label{eq:causal2}\end{equation} if causality is to be preserved.
These quantities can be further simplified by the use of unitarity
relation $S^{\dagger}(x)S(x)=\mathcal{I}$,
expanded similarly in powers of $g(x)$.\\
These are given by\begin{equation}
S_{1}(x)+S_{1}^{\dagger}(x)=0\label{unitary1}\end{equation}
\begin{equation}
S_{2}(x,y)+S_{2}^{\dagger}(x,y)+S_{1}(x)S_{1}^{\dagger}(y)+S_{1}(y)S_{1}^{\dagger}(x)=0\label{unitary2}\end{equation}
 In the case of the local theory, these causality relations {[}(\ref{eq:causal1})
and (\ref{eq:causal2})] are trivially satisfied. In the case of
the nonlocal theories, such quantities, on the other hand, afford
a way of characterizing the causality violation. However, these
quantities contain not the usual S-matrix elements that one can
observe in an experiment (which are obtained with a\textbf{
}\emph{constant i.e. space-time-independent}\textbf{ }coupling),
but rather the coefficients in (\ref{eq:CCexp}). We thus find it
profitable to construct appropriate space-time integrated versions
out of $H_{n}(y,x_{1},...,x_{n})$. Thus, for example, we can
consider \begin{eqnarray}
H_{1} & \equiv & \int d^{4}x\int d^{4}y[\vartheta(x_{0}-y_{0})H_{1}(x,y)+\vartheta(y_{0}-x_{0})H_{1}(y,x)]\nonumber \\
 & = & i\int d^{4}x\int d^{4}yS_{2}(x,y)-i\int d^{4}x\int d^{4}yT[S_{1}(x)S_{1}(y)]\label{eq:H_1}\end{eqnarray}
 which can be expressed entirely in terms of Feynman diagrams that
appear in the usual S-matrix amplitudes. In a similar manner, we
can formulate \begin{eqnarray}
H_{2} & \equiv & \int d^{4}x\int d^{4}y\int d^{4}zH_{2}(x,y,z)\vartheta(x_{0}-y_{0})\vartheta(y_{0}-z_{0})\nonumber \\
 &  & +5\,\mbox{symmetric}\;\mbox{terms}\label{eq:H_2}\end{eqnarray}
 and can itself be expressed in terms of Feynman diagrams.\\
There is a subtle point regarding the expansion (\ref{eq:CCexp})
of the S-matrix in terms of coupling $g$. In a field theory, the
coupling $g$, which has to be the renormalized one, is not a
uniquely defined quantity. In this respect, we have to make a
renormalization convention. In view of the fact that CV, if at all
observed, is expected to be observed at large energies \cite{JJ},
we prefer to use $g$ renormalized at a large energy scale; since
that assures more rapid convergence of the perturbation series. We
shall therefore assume that \[
\widetilde{g}=Re\Gamma^{\left(4\right)}\left(s=-2s_{0}+2m^{2},t=u=s_{0}+m^{2}\right)\]
where $s_{0}$ is a large positive number and $\sqrt{s_{0}}\sim$
C.M. energy of collision. Here, $\Gamma^{\left(4\right)}$ is the
proper 4-point vertex and $s=-2s_{0}+2m^{2},t=u=s_{0}+m^{2}$ is a
point in the unphysical region compatible with $p_{i}^{2}=m^{2}$.
This is equivalent to the following convention: \[
Re\Gamma_{\left(n\right)}^{\left(4\right)}\left(s=-2s_{0}+2m^{2},t=u=s_{0}+m^{2}\right)=0;\qquad
n=1,2,3,...\] where $\Gamma_{\left(n\right)}^{\left(4\right)}$
refers to the $n-$loop contribution to $\Gamma^{\left(4\right)}$. The numerical value of $\tilde{g}$ can be determined by comparing the total experimental cross-section with the expression for it upto a desired order.
\subsection{Results of \cite{JJ} about CV}
In the reference \cite{JJ}, CV in a nonlocal scalar $\phi^4$ theory was studied. It was shown that one can construct amplitudes, which if non-zero, necessarily imply CV. These amplitudes,( $H_1$, $H_2$ etc of \ref{eq:H_1} and \ref{eq:H_2}) can moreover be calculated by means of Feynman diagrams. In \cite{JJ}, causality violation in two exclusive processes (i) $\phi\;\phi\rightarrow \phi\;\phi$ and (ii) $\phi\;\phi\rightarrow \phi\;\phi\;\phi\;\phi$ were studied. It was in particular demonstrated that CV grows significantly with $s$. Here, we shall recall only the result for the first process: $\phi\;\phi\rightarrow \phi\;\phi$  As shown in \cite{JJ}, the s-channel diagram for the CV
amplitude (in the massless) limit yields (the relevant figure, fig. 1, is found in a future section) the following
contribution to the transition amplitude: \[
\Gamma(s)=\frac{{9{\tilde{g}}^{2}}}{{4\pi^{2}}}\sum\limits
_{n=0}^{\infty}{\frac{{\left({\frac{s}{{\Lambda^{2}}}}\right)^{n}\left({1-\frac{1}{{2^{n}}}}\right)}}{{n((n+1)!)}}}\]
 The net causality violating amplitude, considering all the three
$s,t,u$ channels, takes the following form in the massless
limit:\begin{eqnarray}
 &  & \Delta M_{nonlocal}(s,t,u)\\
 & = & \frac{{9{\tilde{g}}^{2}}}{{4\pi^{2}}}\sum\limits _{n=0}^{\infty}{\frac{{\left({1-\frac{1}{{2^{n}}}}\right)}}{{n((n+1)!)}}}\left\{ {\left({\frac{s}{{\Lambda^{2}}}}\right)^{n}+\left({\frac{t}{{\Lambda^{2}}}}\right)^{n}+\left({\frac{u}{{\Lambda^{2}}}}\right)^{n}}\right\}\label{eq:JJresult} \end{eqnarray}
  This CV amplitude is analytic in $s,t,u$ and $m$.
\section{\textsc{\normalsize comparison of cv and local contributions}}
We shall compare the CV terms of (\ref{eq:JJresult}) of \cite{JJ} with the usual
\emph{local} amplitude to get a judgment as to how and when the
former can be isolated.
\subsection{\textsc{\normalsize local theory}}
For the local theory, we find\cite{PS}%
\[ M_{local}=-\frac{{36{\tilde{g}}^{2}}}{{32\pi^{2}}}\left[{\ln s+\ln
t+\ln
u}+constant\right]=-\frac{{36{\tilde{g}}^{2}}}{{32\pi^{2}}}\ln\left[stu\right]+constant\]
In the center of mass frame the Mandelstam variables are given as
follows: \begin{eqnarray*}
s & = & (k_{1}+k_{2})^{2}=(p_{1}+p_{2})^{2}=4p^{2}+4m^{2}\\
t & = & (k_{1}-p_{1})^{2}=(k_{2}-p_{2})^{2}=-2p^{2}(1-cos\theta)\\
u & = &
(k_{1}-p_{2})^{2}=(k_{2}-p_{1})^{2}=-2p^{2}(1+cos\theta)\end{eqnarray*}
So that \begin{eqnarray*}
M_{local} & = & -\frac{{36{\tilde{g}}^{2}}}{{32\pi^{2}}}\left[{\ln stu}\right]+constant\\
 & \approx & -\frac{{36{\tilde{g}}^{2}}}{{32\pi^{2}}}\left[{\ln\left(16p^{6}(1-\cos^{2}\theta)\right)}\right]+constant\end{eqnarray*}
(We have ignored $m^{2}$ compared to $s$ at high energies). The
amplitude can now be expressed in term of the Legendre polynomials
as follows: \begin{eqnarray*}
M_{local} & \equiv & M_{local}(\cos\theta)\\
 & = & \sum\limits _{l=0}^{\infty}{a_{l}^{local}}P_{l}(\cos\theta)\end{eqnarray*}
 Where, \begin{eqnarray*}
a_{l}^{local} & = & \frac{{2l+1}}{2}\int\limits _{-1}^{+1}{M_{local}(\cos\theta)}P_{l}(\cos\theta)d\cos\theta\\
 & = & (-1)^{n}\frac{{2l+1}}{{2^{l+1}l!}}\int\limits _{-1}^{+1}{M_{local}^{n}(x)}\frac{{d^{l-n}}}{{dx^{l-n}}}(x^{2}-1)^{l}dx\\
 & = & (-1)^{n+1}\frac{{2l+1}}{{2^{l+1}l!}}\frac{{36{\tilde{g}}^{2}}}{{32\pi^{2}}}\int\limits _{-1}^{+1}{\frac{{d^{n}}}{{dx^{n}}}[\ln(1-x^{2})]}\frac{{d^{l-n}}}{{dx^{l-n}}}(x^{2}-1)^{l}dx\end{eqnarray*}
 Here $M_{local}^{n}(x)$ stands for the $n$-th derivative of $M_{local}(x)$
with respect to its argument. The coefficients
$a_{2}^{local},a_{4}^{local},a_{6}^{local}$
are obtained%
\myfootnote{The above integrand has a singularity at $x=\pm1$.
This singularity is artificial and presence of $m\neq0$ protects
it. It may appear that setting $m\neq0$ could significantly affect
the values of $a_{l}^{local}$. It has been checked that it is not
the case: In fact $a_{l}^{local}$
are analytic in $m$.%
} as follows: \begin{eqnarray*}
a_{2}^{local} & = & (-1)^{1+1}\frac{5}{{2^{3}2!}}\frac{{36{\tilde{g}}^{2}}}{{32\pi^{2}}}\int\limits _{-1}^{+1}{\frac{d}{{dx}}[\ln(1-x^{2})]}\frac{d}{{dx}}(x^{2}-1)^{2}dx=\frac{{36{\tilde{g}}^{2}}}{{32\pi^{2}}}\left({\frac{5}{3}}\right)\\
a_{4}^{local} & = & (-1)^{1+1}\frac{9}{{2^{5}4!}}\frac{{36{\tilde{g}}^{2}}}{{32\pi^{2}}}\int\limits _{-1}^{+1}{\frac{d}{{dx}}[\ln(1-x^{2})]}\frac{{d^{3}}}{{dx^{3}}}(x^{2}-1)^{4}dx=\frac{{36{\tilde{g}}^{2}}}{{32\pi^{2}}}\left({\frac{9}{{10}}}\right)\\
a_{6}^{local} & = &
(-1)^{3+1}\frac{{13}}{{2^{7}6!}}\frac{{36{\tilde{g}}^{2}}}{{32\pi^{2}}}\int\limits
_{-1}^{+1}{\frac{{d^{3}}}{{dx^{3}}}[\ln(1-x^{2})]}\frac{{d^{3}}}{{dx^{3}}}(x^{2}-1)^{6}dx=\frac{{36{\tilde{g}}^{2}}}{{32\pi^{2}}}\left({\frac{{13}}{{21}}}\right)\end{eqnarray*}
 Therefore, we have \begin{eqnarray}
M_{local} & = & \sum\limits _{l=0}^{\infty}{a_{l}^{local}}P_{l}(\cos\theta)=constant+a_{2}^{local}P_{2}(\cos\theta)+a_{4}^{local}P_{4}(\cos\theta)+a_{6}^{local}P_{6}(\cos\theta)+...\nonumber \\
 & = & \frac{{36{\tilde{g}}^{2}}}{{32\pi^{2}}}\left(constant'+\frac{5}{3}P_{2}(\cos\theta)+\frac{9}{{10}}P_{4}(\cos\theta)+\frac{{13}}{{21}}P_{6}(\cos\theta)+....\right)\label{eq:Pexp}\end{eqnarray}
\subsection{\textsc{\normalsize nonlocal theory~:~$\phi\phi\to\phi\phi$}}
As stated earlier, we wish to compare the CV amplitude of
\cite{JJ} with the local amplitude to see how the former can be
isolated. As shown in \cite{JJ}, the s-channel diagram for the CV
amplitude (in the massless) limit yields the following
contribution to the transition amplitude:
\begin{figure}[!htbp]
\centering
    \includegraphics[scale=0.6]{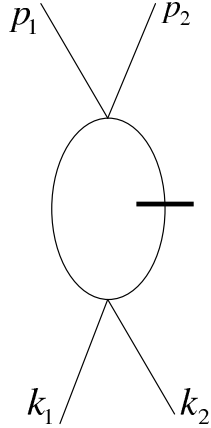}
  \caption{The s-channel diagram giving rise to the one loop causality
violating amplitude $H_{1}$.}
  \label{fig:jI}
\end{figure}
 \[
\Gamma(s)=\frac{{9{\tilde{g}}^{2}}}{{4\pi^{2}}}\sum\limits
_{n=0}^{\infty}{\frac{{\left({\frac{s}{{\Lambda^{2}}}}\right)^{n}\left({1-\frac{1}{{2^{n}}}}\right)}}{{n((n+1)!)}}}\]
 The net causality violating amplitude, considering all the three
$s,t,u$ channels, takes the following form in the massless
limit:\begin{eqnarray*}
 &  & \Delta M_{nonlocal}(s,t,u)\\
 & = & \frac{{9{\tilde{g}}^{2}}}{{4\pi^{2}}}\sum\limits _{n=0}^{\infty}{\frac{{\left({1-\frac{1}{{2^{n}}}}\right)}}{{n((n+1)!)}}}\left\{ {\left({\frac{s}{{\Lambda^{2}}}}\right)^{n}+\left({\frac{t}{{\Lambda^{2}}}}\right)^{n}+\left({\frac{u}{{\Lambda^{2}}}}\right)^{n}}\right\} \end{eqnarray*}
  In the center of mass frame, we have\begin{eqnarray*}
\Delta M_{nonlocal} & = & \frac{{9{\tilde{g}}^{2}}}{{4\pi^{2}}}\sum\limits _{n=0}^{\infty}{\frac{{\left({1-\frac{1}{{2^{n}}}}\right)}}{{n((n+1)!)}}\frac{1}{{\Lambda^{2n}}}}\left({(4p^{2})^{n}+(-2p^{2})^{n}\{(1-\cos\theta)^{n}+(1+\cos\theta)^{n}\}}\right)\\
 & = & \sum_{l=0}{a_{l}^{nonlocal}}P_{l}(\cos\theta)\end{eqnarray*}
Comparison%
\myfootnote{Comparison of \emph{amplitudes }is more natural here,
since the leading contribution from 1-loop calculation depends on
the interference term
which is linear in $M_{local}$ or $\Delta M_{nonlocal}$%
} of $M_{local}$ and $\Delta M_{nonlocal}$ is facilitated by
comparing Legendre coefficients of the same orders. The
coefficients $a_{2}^{nonlocal},a_{4}^{nonlocal}$ and
$a_{6}^{nonlocal}$ are computed as below: \begin{eqnarray*}
a_{2}^{nonlocal} & = & \frac{5}{2}\int\limits _{-1}^{+1}{\Delta M_{nonlocal}^{}(\cos\theta)}P_{2}(\cos\theta)d\cos\theta=\frac{{9{\tilde{g}}^{2}}}{{4\pi^{2}}}\left({\frac{{p^{2}}}{{\Lambda^{2}}}}\right)^{2}\frac{1}{3}\\
a_{4}^{nonlocal} & = & \frac{9}{2}\int\limits _{-1}^{+1}{\Delta M_{nonlocal}^{}(\cos\theta)}P_{4}(\cos\theta)d\cos\theta=\frac{{9{\tilde{g}}^{2}}}{{4\pi^{2}}}\left({\frac{{p^{2}}}{{\Lambda^{2}}}}\right)^{4}\frac{1}{{70}}\\
a_{6}^{nonlocal} & = & \frac{{13}}{2}\int\limits _{-1}^{+1}{\Delta
M_{nonlocal}^{}(\cos\theta)}P_{6}(\cos\theta)d\cos\theta=\frac{{9{\tilde{g}}^{2}}}{{4\pi^{2}}}\left({\frac{{p^{2}}}{{\Lambda^{2}}}}\right)^{6}\frac{1}{{3465}}\end{eqnarray*}
\newpage
\begin{table}
We summarize the ratio of the local and nonlocal coefficients and
their numerical values in the following table:
\begin{center}
\begin{tabular}{|c|c|c|c||c|}
\hline Ratio of coefficients & $\frac{p^{2}}{\Lambda^{2}}=0.1$ &
$\frac{p^{2}}{\Lambda^{2}}=0.2$ &
\multicolumn{1}{c||}{$\frac{p^{2}}{\Lambda^{2}}=0.4$} &
$\frac{p^{2}}{\Lambda^{2}}=0.8$\tabularnewline \hline \hline
$\frac{a_{2}^{nonlocal}}{a_{2}^{local}}$ & 0.4\% & 1.6\% & 6.4\% &
25.6\%\myfootnotemark \tabularnewline \hline
$\frac{a_{4}^{nonlocal}}{a_{4}^{local}}$ & 0.00032\% & 0.0051\% &
0.081\% & 1.3\%\tabularnewline \hline
$\frac{a_{6}^{nonlocal}}{a_{6}^{local}}$ & 9.3$\times10^{-10}$ &
6$\times10^{-8}$ & 3.8$\times10^{-6}$ &
2.4$\times10^{-4}$\tabularnewline \hline
\end{tabular}
\caption{Comparison of local and non-local contributions for
coefficients of different Legendre polynomials.}
 \label{T:dimens}
 \end{center}
 These ratios are independent of the coupling constant $g$. It
appears that there is a significant chance of detecting CV only in
the ratio $\frac{a_{2}^{nonlocal}}{a_{2}^{local}}$ and when
$p^{2}\lesssim\Lambda^{2}$.\\
Finally we point out that while we have picked up the process $\phi\phi\rightarrow\phi\phi$ for simplicity, this would not be the process for which observation of CV is the most efficient. This is so because as pointed out in \cite{JJ}, the CV in this process is of higher order in $\frac{p^{2}}{\Lambda^{2}}$, viz. $O\left(\frac{p^{4}}{\Lambda^{4}}\right)$. CV should be more noticeable in a process such as $\phi\phi\rightarrow\phi\phi\phi\phi$.
\end{table}
\myfootnotetext{When the ratio is large, higher order corrections
to CV cannot be ignored.}
\section{\textsc{\normalsize construction of observables\label{sec:4}}}
In this section, we shall construct a quantity, partly dependent
on physically observable differential cross-section and partly on
perturbative calculations, which can detect CV. Of course, we make
use of the quantities $H_{1}$ of eq.(\ref{eq:H_1}) which signal CV
\cite{JJ,J06}. The S-operator has the expansion\footnote{Henceforth, we have often suppressed "tilde" on $g$}: \[ S=1+g\int
d^{4}xS_{1}\left(x\right)+\frac{g^{2}}{2!}\int
d^{4}xd^{4}yS_{2}\left(x,y\right)+..........\] Consider a
following matrix element between some initial and final states
$\left|i\right\rangle $ and $\left|f\right\rangle $:\[
\left\langle f\right|S\left|i\right\rangle =\delta_{fi}+g\int
d^{4}x\left\langle f\right|S_{1}\left(x\right)\left|i\right\rangle
+\frac{g^{2}}{2!}\int d^{4}xd^{4}y\left\langle
f\right|S_{2}\left(x,y\right)\left|i\right\rangle +..........\] We
have, from translational invariance,\[ \int d^{4}x\left\langle
f\right|S_{1}\left(x\right)\left|i\right\rangle =\int
d^{4}x\left\langle f\right|S_{1}\left(0\right)\left|i\right\rangle
e^{i\left(p_{f}-p_{i}\right).x}=\left\langle
f\right|S_{1}\left(0\right)\left|i\right\rangle
\left(2\pi\right)^{4}\delta^{4}\left(p_{f}-p_{i}\right)\]
Expressing $x=\left(\xi+\eta\right)/2$ and
$y=\left(\eta-\xi\right)/2$, we have
\begin{eqnarray*}
\int{d^{4}xd^{4}y}\left\langle {f\left|{S_{2}(x,y)}\right|i}\right\rangle  & = & \left(\frac{1}{2}\right)^{4}\int{d^{4}\xi d^{4}\eta}\left\langle {f\left|{S_{2}\left(\frac{{\eta+\xi}}{2},\frac{{\eta-\xi}}{2}\right)}\right|i}\right\rangle \\
 & = & \left(\frac{1}{2}\right)^{4}\int{d^{4}\xi d^{4}\eta}\left\langle {f\left|{e^{iP.\frac{\eta}{2}}S_{2}\left(\frac{\xi}{2},-\frac{\xi}{2}\right)e^{-iP.\frac{\eta}{2}}}\right|i}\right\rangle \\
 & = & \left(\frac{1}{2}\right)^{4}\int{d^{4}\xi d^{4}\eta}\left\langle {f\left|{e^{ip_{f}.\frac{\eta}{2}}S_{2}\left(\frac{\xi}{2},-\frac{\xi}{2}\right)e^{-ip_{i}.\frac{\eta}{2}}}\right|i}\right\rangle \\
 & = & \int{d^{4}\xi}\left\langle {f\left|S_{2}\left(\frac{\xi}{2},-\frac{\xi}{2}\right)\right|i}\right\rangle (2\pi)^{4}\delta^{4}(p_{f}-p_{i})\end{eqnarray*}
The S-matrix is related to the invariant matrix element
$\mathcal{M}_{fi}$ as:\begin{eqnarray*}
{\left\langle f\left|S\right|i\right\rangle } & \equiv S_{fi} & =\left\langle {f|i}\right\rangle +i\left\langle {f\left|T\right|i}\right\rangle \\
 & = & \left\langle {f|i}\right\rangle +i(2\pi)^{4}\delta^{4}(p_{f}-p_{i})\mathcal{M}_{fi}\end{eqnarray*}
Thus,\begin{eqnarray*}
\mathcal{M}_{fi} & = & -ig\left\langle {f\left|{S_{1}(0)}\right|i}\right\rangle +\frac{{-ig^{2}}}{2}\int{d^{4}\xi}\left\langle {f\left|S_{2}\left(\frac{\xi}{2},-\frac{\xi}{2}\right)\right|i}\right\rangle \\
 & + & \frac{-ig^{3}}{3!}\int d^{4}\xi d^{4}\eta\left\langle {f\left|S_{3}\left(0,\xi,\eta\right)\right|i}\right\rangle +.....\\
 & \equiv & \mathcal{M}^{(1)}+\mathcal{M}^{(2)}+\mathcal{M^{\mathit{\mathrm{(3)}}}}+....\end{eqnarray*}
Now, consider the exclusive scattering process:
$\phi\left(k_{1}\right)+\phi\left(k_{2}\right)\rightarrow\phi\left(p_{1}\right)+\phi\left(p_{2}\right)$.
The differential cross-section reads: \[
\frac{{d\sigma}}{{d^{3}p_{1}d^{3}p_{2}}}=\frac{1}{2}\frac{{(2\pi)^{4}\delta^{4}(p_{f}-p_{i})}}{{2\omega_{p_{1}}2\omega_{p_{2}}\left|{\vec{v}_{1}-\vec{v}_{2}}\right|2\omega_{k_{1}}2\omega_{k_{2}}}}\left|{\mathcal{M}}\right|^{2}\]
Here, $p_{i}=k_{1}+k_{2}$ and $p_{f}=p_{1}+p_{2}$ and
$\mathbf{v_{1,2}}$ are velocities of the colliding particles and
$\frac{1}{2}$ is the symmetry factor. {[}We are using the
conventions as outlined in \cite{PS}]. We integrate over $p_{2}$
using the $\delta^{3}\left(\mathbf{p_{f}-p_{i}}\right)$. We
express $d^{3}p_{1}=p_{1}^{2}dp_{1}d\Omega$, integrate over
$p_{1}$ to find,\[ \frac{{d\sigma}}{{d\Omega}}=\int
p_{1}^{2}dp_{1}\frac{1}{2}\frac{{(2\pi)^{4}\delta(k_{10}+k_{20}-p_{10}-p_{20})}}{{2\omega_{p_{1}}2\omega_{p_{2}}\left|{\vec{v}_{1}-\vec{v}_{2}}\right|2\omega_{k_{1}}2\omega_{k_{2}}}}\left|{\mathcal{M}}\right|^{2}\]
 In the C.M. frame, $k_{10}+k_{20}\equiv2\sqrt{k^{2}+m^{2}}\equiv2\omega_{k}$
and $p_{10}+p_{20}\equiv2\sqrt{p^{2}+m^{2}}=2\omega_{p}$. So
that,\[
\frac{{d\sigma}}{{d\Omega}}=\frac{p\omega_{p}}{4}\frac{{(2\pi)^{4}}}{{2\omega_{p_{1}}2\omega_{p_{2}}\left|{\vec{v}_{1}-\vec{v}_{2}}\right|2\omega_{k_{1}}2\omega_{k_{2}}}}\left|{\mathcal{M}}\right|^{2}\]
Here, \begin{eqnarray*}
\left|\mathcal{M}\right|^{2} & = & \left|{-ig\left\langle {f\left|{S_{1}(0)}\right|i}\right\rangle -i\frac{{g^{2}}}{2}\int{d\xi}\left\langle {f\left|S_{2}\left(\frac{\xi}{2},-\frac{\xi}{2}\right)\right|i}\right\rangle +....}\right|^{2}\\
 & = & g^{2}\left|{\left\langle {f\left|{S_{1}(0)}\right|i}\right\rangle }\right|^{2}+g^{3}{\rm Re}\left[{\left\langle {f\left|{S_{1}(0)}\right|i}\right\rangle ^{*}\int{d\xi}\left\langle {f\left|S_{2}\left(\frac{\xi}{2},-\frac{\xi}{2}\right)\right|i}\right\rangle }\right]+\mathcal{R}\end{eqnarray*}
$\mathcal{R}$ are the $O\left(g^{4}\right)$ terms:\begin{eqnarray}
\mathcal{R} & \equiv & \frac{g^{4}}{4}\left|\int{d\xi}\left\langle {f\left|S_{2}\left(\frac{\xi}{2},-\frac{\xi}{2}\right)\right|i}\right\rangle \right|^{2}\nonumber \\
 & + & 2\frac{g^{4}}{3!}Re\left[\left\langle {f\left|{S_{1}(0)}\right|i}\right\rangle ^{*}\int d^{4}\xi d^{4}\eta\left\langle {f\left|S_{3}\left(0,\xi,\eta\right)\right|i}\right\rangle \right]\label{eq:aar}\end{eqnarray}
The differential cross-section, now becomes: \begin{eqnarray*}
\frac{{d\sigma}}{{d\Omega}} & = & \frac{p\omega_{p}}{4}(2\pi)^{4}\\
 & \times & \frac{{\left\{ {g^{2}\left|{\left\langle {f\left|{S_{1}(0)}\right|i}\right\rangle }\right|^{2}+g^{3}{\rm Re}\left[{\left\langle {f\left|{S_{1}(0)}\right|i}\right\rangle ^{*}\int{d\xi}\left\langle {f\left|S_{2}\left(\frac{\xi}{2},-\frac{\xi}{2}\right)\right|i}\right\rangle }\right]+O(g^{4})}\right\} }}{{2\omega_{p_{1}}2\omega_{p_{2}}\left|{\vec{v}_{1}-\vec{v}_{2}}\right|2\omega_{k_{1}}2\omega_{k_{2}}}}\end{eqnarray*}
Now,\begin{eqnarray*} iM & = & g\left\langle
{p_{1}p_{2}\left|{S_{1}(0)}\right|k_{1}k_{2}}\right\rangle
\end{eqnarray*}
 is the lowest order amplitude which equals $-6ig$. Therefore it
is required that \[ \left\langle
{p_{1}p_{2}\left|{S_{1}(0)}\right|k_{1}k_{2}}\right\rangle =-6i\]
Thus,\begin{eqnarray*}
 &  & \frac{{d\sigma}}{{d\Omega}}\\
 & = & \frac{p\omega_{p}}{4}\frac{{(2\pi)^{4}\left\{ {36g^{2}-6g^{3}Im\left[\int{d\xi}\left\langle {f\left|S_{2}\left(\frac{\xi}{2},-\frac{\xi}{2}\right)\right|i}\right\rangle \right]+O(g^{4})}\right\} }}{{2\omega_{p_{1}}2\omega_{p_{2}}\left|{\vec{v}_{1}-\vec{v}_{2}}\right|2\omega_{k_{1}}2\omega_{k_{2}}}}\end{eqnarray*}
and subtracting the angular average of
$\frac{{d\sigma}}{{d\Omega}}$\begin{eqnarray}
 &  & \frac{{d\sigma}}{{d\Omega}}-\overline{\frac{{d\sigma}}{{d\Omega}}}\nonumber \\
 & = & \frac{p\omega_{p}}{4}(2\pi)^{4}\nonumber \\
 & \times & \frac{{\left\{ {-6g^{3}Im\left[\int{d\xi}\left\langle {f\left|S_{2}\left(\frac{\xi}{2},-\frac{\xi}{2}\right)\right|i}\right\rangle -\overline{\int{d\xi}\left\langle {f\left|S_{2}\left(\frac{\xi}{2},-\frac{\xi}{2}\right)\right|i}\right\rangle }\right]+O(g^{4})}\right\} }}{{2\omega_{p_{1}}2\omega_{p_{2}}\left|{\vec{v}_{1}-\vec{v}_{2}}\right|2\omega_{k_{1}}2\omega_{k_{2}}}}
\label{eq:sigma-}
\end{eqnarray} Consider the following matrix
element of $H_{1}$ of eq.(\ref{eq:H_1}):\begin{eqnarray*}
\left\langle {p_{1}p_{2}\left|{H_{1}}\right|k_{1}k_{2}}\right\rangle  & = & i\int{d^{4}xd^{4}y\left\{ \left\langle {p_{1}p_{2}\left|{S_{2}(x,y)}\right|k_{1}k_{2}}\right\rangle -\left\langle {p_{1}p_{2}\left|{T[S_{1}(x)S_{1}(y)]}\right|k_{1}k_{2}}\right\rangle \right\} }\\
 & = & (2\pi)^{4}\delta^{4}(p_{f}-k_{i})i\left\{ \int{d\xi}\left\langle {p_{1}p_{2}\left|{S_{2}(\frac{\xi}{2},-\frac{\xi}{2})}\right|k_{1}k_{2}}\right\rangle \right.\\
 & - & \left.\int{d\xi}\left\langle {p_{1}p_{2}\left|{T[S_{1}(\frac{\xi}{2})S_{1}(-\frac{\xi}{2})]}\right|k_{1}k_{2}}\right\rangle \right\} \end{eqnarray*}
We now integrate over $\mathbf{p_{2}}$ followed by
$p_{1}=|\mathbf{p_{1}}|$ as before to obtain,\begin{eqnarray*}
\int p_{1}^{2}dp_{1}Re\left\langle {p_{1}p_{2}\left|{H_{1}}\right|k_{1}k_{2}}\right\rangle  & = & -(2\pi)^{4}\frac{p\omega_{p}}{2}Im\left\{ \int{d\xi}\left\langle {p_{1}p_{2}\left|{S_{2}(\frac{\xi}{2},-\frac{\xi}{2})}\right|k_{1}k_{2}}\right\rangle \right.\\
 & - & \left.\int{d\xi}\left\langle {p_{1}p_{2}\left|{T[S_{1}(\frac{\xi}{2})S_{1}(-\frac{\xi}{2})]}\right|k_{1}k_{2}}\right\rangle \right\} \end{eqnarray*}
where we set $\mathbf{p_{2}=k_{1}+k_{2}-p_{1}}$. Left hand side is
a function of angular variables: $\Omega$. We subtract out the
angular average to find,
\begin{eqnarray*}
 &  & \int p_{1}^{2}dp_{1}Re\left\{ \left\langle {p_{1}p_{2}\left|{H_{1}}\right|k_{1}k_{2}}\right\rangle -\overline{\left\langle {p_{1}p_{2}\left|{H_{1}}\right|k_{1}k_{2}}\right\rangle }\right\} \\
 & = & -(2\pi)^{4}\frac{p\omega_{p}}{2}Im\left\{ {\int{d\xi}\left\langle {p_{1}p_{2}\left|{S_{2}\left(\frac{\xi}{2},-\frac{\xi}{2}\right)}\right|k_{1}k_{2}}\right\rangle }\right.\\
 & - & \left.\overline{\int{d\xi}\left\langle {p_{1}p_{2}\left|{S_{2}(\frac{\xi}{2},-\frac{\xi}{2})}\right|k_{1}k_{2}}\right\rangle }\right\} \\
 & + & (2\pi)^{4}\frac{p\omega_{p}}{2}Im\left\{ \int{d\xi}\left\langle {p_{1}p_{2}\left|{T[S_{1}(\frac{\xi}{2})S_{1}(-\frac{\xi}{2})]}\right|k_{1}k_{2}}\right\rangle \right.\\
 & - & \left.\overline{\int{d\xi}\left\langle {p_{1}p_{2}\left|{T[S_{1}(\frac{\xi}{2})S_{1}(-\frac{\xi}{2})]}\right|k_{1}k_{2}}\right\rangle }\right\} \end{eqnarray*}
Now we employ (\ref{eq:sigma-}) to obtain,
\begin{eqnarray}
 &  & \int p_{1}^{2}dp_{1}Re\left\{ \left\langle {p_{1}p_{2}\left|{H_{1}}\right|k_{1}k_{2}}\right\rangle -\overline{\left\langle {p_{1}p_{2}\left|{H_{1}}\right|k_{1}k_{2}}\right\rangle }\right\} \nonumber \\
 & = & \frac{1}{3g^{3}}{2\omega_{p_{1}}2\omega_{p_{2}}\left|{\vec{v}_{1}-\vec{v}_{2}}\right|2\omega_{k_{1}}2\omega_{k_{2}}}\left[\frac{{d\sigma}}{{d\Omega}}-\overline{\frac{{d\sigma}}{{d\Omega}}}+O\left(g^{4}\right)\right]\nonumber \\
 & + & (2\pi)^{4}\frac{p\omega_{p}}{2}Im\left\{ \int{d\xi}\left\langle {p_{1}p_{2}\left|{T[S_{1}(\frac{\xi}{2})S_{1}(-\frac{\xi}{2})]}\right|k_{1}k_{2}}\right\rangle \right.\nonumber \\
 & - & \left.\overline{\int{d\xi}\left\langle {p_{1}p_{2}\left|{T[S_{1}(\frac{\xi}{2})S_{1}(-\frac{\xi}{2})]}\right|k_{1}k_{2}}\right\rangle }\right\} \nonumber \\
 & = & \frac{32\omega^{3}p}{3g^{3}}\left[\frac{{d\sigma}}{{d\Omega}}-\overline{\frac{{d\sigma}}{{d\Omega}}}+O\left(g^{4}\right)\right]\nonumber \\
 & + & (2\pi)^{4}\frac{p\omega_{p}}{2}Im\left\{ \int{d\xi}\left\langle {p_{1}p_{2}\left|{T[S_{1}(\frac{\xi}{2})S_{1}(-\frac{\xi}{2})]}\right|k_{1}k_{2}}\right\rangle \right.\nonumber \\
 &  & -\left.\overline{\int{d\xi}\left\langle {p_{1}p_{2}\left|{T[S_{1}(\frac{\xi}{2})S_{1}(-\frac{\xi}{2})]}\right|k_{1}k_{2}}\right\rangle }\right\} \label{eq:CVO}\end{eqnarray}
Causality necessarily requires that the left hand side of
(\ref{eq:CVO}) vanishes. On the right hand side, there are
\begin{enumerate}
\item experimentally observable quantity,
$\frac{{d\sigma}}{{d\Omega}}-\overline{\frac{{d\sigma}}{{d\Omega}}}$,
\item a theoretically calculable quantity (by a Feynman diagram calculation)\begin{eqnarray*}
 &  & Im\left\{ \int{d^{4}\xi}\left\langle {p_{1}p_{2}\left|{T[S_{1}(\frac{\xi}{2})S_{1}(-\frac{\xi}{2})]}\right|k_{1}k_{2}}\right\rangle \right.\\
 & - & \left.\overline{\int{d^{4}\xi}\left\langle {p_{1}p_{2}\left|{T[S_{1}(\frac{\xi}{2})S_{1}(-\frac{\xi}{2})]}\right|k_{1}k_{2}}\right\rangle }\right\} \end{eqnarray*}
and
\item $O\left(g^{4}\right)$ and higher order terms from $\frac{{d\sigma}}{{d\Omega}}-\overline{\frac{{d\sigma}}{{d\Omega}}}$
in addition to\[ Im\left\{ {\int{d^{4}\xi}\left\langle
{p_{1}p_{2}\left|{S_{2}\left(\frac{\xi}{2},-\frac{\xi}{2}\right)}\right|k_{1}k_{2}}\right\rangle
-\overline{\int{d^{4}\xi}\left\langle
{p_{1}p_{2}\left|{S_{2}\left(\frac{\xi}{2},-\frac{\xi}{2}\right)}\right|k_{1}k_{2}}\right\rangle
}}\right\}.\]
\end{enumerate}
We shall calculate the second quantity in the coming section. We
shall also explain how and when the $O\left(g^{4}\right)$ term can
be ignored.
\section{{\normalsize \textsc{contribution of the second term in} (\ref{eq:CVO})}}
As seen in \cite{JJ}, the second term in (\ref{eq:CVO})
corresponds to the fish diagram with smeared propagators shown
below. It is calculated in the massless limit below:
\begin{figure}[!htbp]
\centering
    \includegraphics[scale=.6]{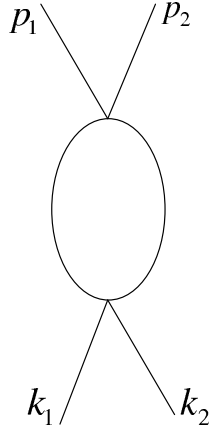}
  \caption{The Feynman diagram equivalent to the second term
$\int{d\xi}\left\langle
{p_{1}p_{2}\left|{T[S_{1}(\frac{\xi}{2})S_{1}(-\frac{\xi}{2})]}\right|k_{1}k_{2}}\right\rangle
$ . Only the s-channel diagram is shown.}
  \label{fig:jI}
\end{figure}
One finds,
\[
\Gamma_{s}=\frac{{9g^{2}}}{{8\pi^{2}}}\int\limits
_{1}^{\infty}{d\tau_{1}\int\limits
_{1}^{\infty}{d\tau_{2}}}\frac{{e^{-\frac{{(P)^{2}}}{{\Lambda^{2}}}\frac{{\tau_{1}\tau_{2}}}{{\tau_{1}+\tau_{2}}}}}}{{(\tau_{1}+\tau_{2})^{2}}}\]
 where $P^2=-(p_1+p_2)^2=-s$ and is positive in Euclidean space. We employ \cite{KW92}, \[
\int\limits _{1}^{\infty}{d\tau_{1}\int\limits
_{1}^{\infty}{d\tau_{2}}=}\int\limits
_{\frac{1}{2}}^{1}{dx\int\limits _{\frac{1}{{1-x}}}^{\infty}{\tau
d\tau}}+\int\limits _{0}^{\frac{1}{2}}{dx\int\limits
_{\frac{1}{x}}^{\infty}{\tau d\tau}}\] {[}where
$\tau=\tau_{1}+\tau_{2}$ and $x=\frac{\tau_{2}}{\tau}$] and find
\[ \Gamma_{s}=\frac{{9g^{2}}}{{8\pi^{2}}}\left({\int\limits
_{\frac{1}{2}}^{1}{dx\int\limits
_{\frac{1}{{1-x}}}^{\infty}{d\tau+\int\limits
_{0}^{\frac{1}{2}}{dx\int\limits
_{\frac{1}{x}}^{\infty}{d\tau}}}}}\right)\frac{{e^{-\frac{P^2}{{\Lambda^{2}}}\tau(1-x)x}}}{{\tau}}\]
setting $t=\frac{P^2}{{\Lambda^{2}}}\tau(1-x)x$\begin{eqnarray*}
\Gamma_{s} & = & \frac{{9g^{2}}}{{8\pi^{2}}}\left({\int\limits _{\frac{1}{2}}^{1}{dx\int\limits _{\frac{{P^{2}x}}{{\Lambda^{2}}}}^{\infty}{dt+\int\limits _{0}^{\frac{1}{2}}{dx\int\limits _{\frac{{P^2(1-x)}}{{\Lambda^{2}}}}^{\infty}{dt}}}}}\right)\frac{{e^{-t}}}{t}\\
 & = & \frac{{9g^{2}}}{{8\pi^{2}}}\int\limits _{\frac{1}{2}}^{1}{dx\Gamma\left(0,\frac{{P^{2}x}}{{\Lambda^{2}}}\right)+\int\limits _{0}^{\frac{1}{2}}{dx\Gamma\left(0,\frac{{P^2(1-x)}}{{\Lambda^{2}}}\right)}};\\
 & = & \frac{{9g^{2}}}{{4\pi^{2}}}\int\limits _{\frac{1}{2}}^{1}{dx\Gamma(0,\frac{{P^2x}}{{\Lambda^{2}}}){\rm {}}}\\
 & = & \frac{{9g^{2}}}{{4\pi^{2}}}\int\limits _{\frac{1}{2}}^{1}{dx\left[{-\ln\frac{{P^{2}x}}{{\Lambda^{2}}}-\gamma-\sum\limits _{n=1}^{\infty}{\frac{{(-\frac{{P^{2}x}}{{\Lambda^{2}}})^{n}}}{{n(n!)}}}}\right]}\\
 & = & \frac{{9g^{2}}}{{8\pi^{2}}}\left({-\ln\frac{s}{{\Lambda^{2}}}+constant-2\sum\limits _{n=1}^{\infty}{\frac{{1}}{{n(n+1)!}}\left({\frac{s}{{\Lambda^{2}}}}\right)^{n}\left({1-\frac{1}{{2^{n+1}}}}\right)}}\right)\end{eqnarray*}
where, $\Gamma\left(n,z\right)$ is the incomplete
$\Gamma$-function\[ \Gamma(n,z)\equiv\int\limits
_{z}^{\infty}{\frac{{dt}}{t}}t^{n}e^{-t}\] Adding up
$s,t,u$-channels together, \begin{eqnarray*}
\Gamma(s,t,u) & = & \frac{{9g^{2}}}{{8\pi^{2}}}\left[{-\ln\frac{s}{{\Lambda^{2}}}-\ln\frac{t}{{\Lambda^{2}}}-\ln\frac{u}{{\Lambda^{2}}}+constant}\right.\\
 &  & \left.{-2\sum\limits _{n=1}^{\infty}{\frac{{1}}{{\left(n+1\right)(n!)}}\left({\left({\frac{s}{{\Lambda^{2}}}}\right)^{n}+\left({\frac{t}{{\Lambda^{2}}}}\right)^{n}+\left({\frac{u}{{\Lambda^{2}}}}\right)^{n}}\right)\left({1-\frac{1}{{2^{n+1}}}}\right)}}\right]\end{eqnarray*}
As we shall be interested in the nontrivial contribution arising
from nonlocal effects, we shall find it convenient to filter out
the usual local effects. We parametrize the local part of the
above expression as\[
l\left(\theta\right)=c_1+c_2ln\left(1-cos^{2}\theta\right)\] Using
(\ref{eq:Pexp}),\begin{eqnarray*} l\left(\theta\right) & = &
c'_1+c_2\left(\frac{5}{3}P_{2}(\cos\theta)+\frac{9}{{10}}P_{4}(\cos\theta)+\frac{{13}}{{21}}P_{6}(\cos\theta)+....\right)\end{eqnarray*}
So that, the quantity entering in the equation (\ref{eq:CVO}) is
\[
l\left(\theta\right)-\overline{l}\left(\theta\right)=c_2\left(\frac{5}{3}P_{2}(\cos\theta)+\frac{9}{{10}}P_{4}(\cos\theta)+\frac{{13}}{{21}}P_{6}(\cos\theta)+....\right)\]
Now, consider\begin{eqnarray*} h\left(\theta\right) & = & \alpha
P_{2}\left(cos\theta\right)+\beta
P_{4}\left(cos\theta\right)\end{eqnarray*}
$h\left(\theta\right)$ is the simplest non-trivial even polynomial \emph{orthogonal
}to $l\left(\theta\right)-\overline{l}\left(\theta\right)$
provided\[ \frac{2}{3}\alpha+\frac{1}{5}\beta=0\] We choose to
integrate (\ref{eq:CVO}) with $h\left(\theta\right)$. Thus the CV
signaling amplitude may be conveniently as\[ \int dcos\theta
h\left(\theta\right)\left\{
\frac{32\omega^{3}p}{3g^{3}}\left[\frac{{d\sigma}}{{d\Omega}}\right]+(2\pi)^{4}\frac{p\omega_{p}}{2}Im\left\{
\int{d\xi}\left\langle
{p_{1}p_{2}\left|{T[S_{1}(\frac{\xi}{2})S_{1}(-\frac{\xi}{2})]}\right|k_{1}k_{2}}\right\rangle
\right\} \right\} \] where we have dropped the two terms with
angular averages as $\int dcos\theta h\left(\theta\right)\times
constant=0$.
\section{\textsc{\normalsize $O\left(g^{4}\right)$ contributions}}
We shall calculate $O\left(g^{4}\right)$ terms in $\mathcal{R}$ of
equation (\ref{eq:aar}) and find the range of couplings and
energies when it is ignorable. Calculations of quantities required
for this has already been done in a \emph{local }theory. As such
quantities in a nonlocal theory will differ only by terms of
$O\left(\frac{1}{\Lambda^{2}}\right)$ from a local theory and we
are interested only in an estimate of such terms in $\mathcal{R}$,
we shall employ the \emph{local }results for this purpose.
\subsection{\textsc {the $S_{1}S_{3}$-type terms}}
One of the contributions to $\frac{d\sigma}{d\Omega}$ we have not
taken account of is the $O\left(g^{4}\right)$ contribution coming
from a term of the kind $S_{1}S_{3}$ in $\mathcal{R}$. To evaluate
this we need to calculate $O\left(g^{3}\right)$ contribution to
S-matrix coming from the two loop diagrams.
\begin{figure}[!htbp]
\centering
    \includegraphics[scale=0.6]{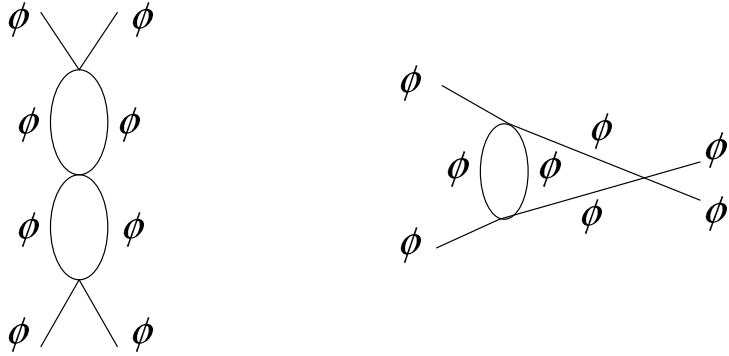}
  \caption{Diagrams contributing to 2 particle matrix element of
$S_{3}$. Diagrams obtained by interchanges of momentum labels are
not shown.}
  \label{fig:jI}
\end{figure}
These two loop diagrams have already been computed in the context
of the standard model \cite{MDR} with the renormalization
convention which amounts to using a mass scale $\sim m$. We shall
adopt the result to the case of $\phi^{4}$ theory and employ them
with our renormalization convention. The leading terms in the
amplitude $A\left(s,t,u\right)$ comes from the $ln^{2}s,\,
ln^{2}t,\, ln^{2}u$ terms for $s$ large. Keeping these terms, and
using the renormalization convention of \cite{MDR}, the full
amplitude $A\left(s,t,u\right)$ is {[}here,
$\hat{s}=\frac{s}{m^{2}}$ etc.]
\begin{eqnarray*}
A(s,t,u) & = & -6g+\frac{{g^{2}}}{{16\pi^{2}}}\left[{-18\left({\ln(-\hat{s})+\ln(-\hat{t})+\ln(-\hat{u})}\right)}\right]\\
 & + & \frac{{g^{3}}}{{\left({16\pi^{2}}\right)^{2}}}\left[{-162\left({\ln^{2}(-\hat{s})+\ln^{2}(-\hat{t})+\ln^{2}(-\hat{u})}\right)+.....}\right]\\
 & \equiv & -6g+g^{2}a+g^{3}b\end{eqnarray*}
 Where, \begin{eqnarray*}
a & = & \frac{1}{{16\pi^{2}}}\left[{-18\left({\ln(\frac{{-s}}{{m^{2}}})+\ln(\frac{{-t}}{{m^{2}}})+\ln(\frac{{-u}}{{m^{2}}})}\right)}\right]\\
b & = &
\frac{1}{{\left({16\pi^{2}}\right)^{2}}}\left[{-162\left({\ln^{2}(\frac{{-s}}{{m^{2}}})+\ln^{2}(\frac{{-t}}{{m^{2}}})+\ln^{2}(\frac{{-u}}{{m^{2}}})}\right)+.....}\right]\end{eqnarray*}
 We define $\tilde{g}$ by evaluating $Re\left[A\left(s,t,u\right)\right]$
at $s=-2s_{0}+2m^{2},\; t=u=s_{0}+m^{2}$. We have,
\begin{equation}
\begin{array}{l}
-6\tilde{g}=-6g+g^{2}\tilde{a}+g^{3}\tilde{b}+O(g^{4})\end{array}\label{eq:ggtilde}\end{equation}
 Where, \begin{eqnarray*}
\tilde{a} & = & \frac{1}{{16\pi^{2}}}\left[{-18\left({\ln(\frac{{2s_{0}}}{{m^{2}}})+2\ln(\frac{{s_{0}}}{{m^{2}}})}\right)}\right]\\
\tilde{b} & = &
\frac{1}{{\left({16\pi^{2}}\right)^{2}}}\left[{-162\left({\ln^{2}(\frac{{2s_{0}}}{{m^{2}}})+2\ln^{2}(\frac{{s_{0}}}{{m^{2}}})}\right)+.....}\right]\end{eqnarray*}
 Now from (\ref{eq:ggtilde}), \begin{eqnarray*}
-6g & = & -6\tilde{g}-g^{2}\tilde{a}-g^{3}\tilde{b}+O(g^{4})\\
 & = & -6\tilde{g}-\left({\tilde{g}+\frac{{g^{2}}}{6}\tilde{a}+..}\right)^{2}\tilde{a}-\left({\tilde{g}+\frac{{g^{2}}}{6}\tilde{a}+..}\right)^{3}\tilde{b}+O(\tilde{g}^{4})\\
 & = & -6\tilde{g}-\tilde{g}^{2}\tilde{a}-\tilde{g}^{3}\left({\frac{{\tilde{a}^{2}}}{3}+\tilde{b}}\right)+O(\tilde{g}^{4})\end{eqnarray*}
 So that we can express $A\left(s,t,u\right)$ in terms of $\tilde{g}$
as, \begin{eqnarray*}
A(s,t,u) & = & -6g+g^{2}a+g^{3}b\\
 & = & -6\tilde{g}-\tilde{g}^{2}\tilde{a}-\tilde{g}^{3}\left({\frac{{\tilde{a}^{2}}}{3}+\tilde{b}}\right)+\tilde{g}^{2}a+\tilde{g}^{3}b+\frac{{\tilde{g}^{3}}}{3}\tilde{a}a+o(\tilde{g}^{4})\\
 & = & -6\tilde{g}-\tilde{g}^{2}(\tilde{a}-a)-\tilde{g}^{3}\left({\frac{{\tilde{a}^{2}}}{3}+\tilde{b}-b-\frac{{\tilde{a}a}}{3}}\right)+o(\tilde{g}^{4})\end{eqnarray*}
where,
\[
\tilde{a}-a=\frac{1}{{16\pi^{2}}}\left[{-18\left\{
{\ln\left({\frac{{2s_{0}}}{s}}\right)+\ln\left({\frac{{s_{0}}}{t}}\right)+\ln\left({\frac{{s_{0}}}{u}}\right)}\right\}
}\right]\] To calculate the relevant matrix element of $S_{3}$, we
need to focus our attention on the coefficient of
$\tilde{g}^{3}$\begin{eqnarray*}
 &  & \frac{{\tilde{a}^{2}}}{3}+\tilde{b}-b-\frac{{\tilde{a}a}}{3}\\
 & = & \frac{1}{{\left({16\pi^{2}}\right)^{2}}}\left\{ {108\left[{\left({\ln(\frac{{2s_{0}}}{{m^{2}}})+2\ln(\frac{{s_{0}}}{{m^{2}}})}\right)^{2}}\right.}\right.\\
 & - & \left.{\left({\ln(\frac{{2s_{0}}}{{m^{2}}})+2\ln(\frac{{s_{0}}}{{m^{2}}})}\right)\left({\ln(\frac{{-s}}{{m^{2}}})+\ln(\frac{{-t}}{{m^{2}}})+\ln(\frac{{-u}}{{m^{2}}})}\right)}\right]\\
 & - & \left.{162\left({\ln^{2}(\frac{{2s_{0}}}{{m^{2}}})+\ln^{2}(\frac{{s_{0}}}{{m^{2}}})+\ln^{2}(\frac{{s_{0}}}{{m^{2}}})-\ln^{2}(\frac{{-s}}{{m^{2}}})-\ln^{2}(\frac{{-t}}{{m^{2}}})-\ln^{2}(\frac{{-u}}{{m^{2}}})}\right)}\right\} \\
 & = & \frac{1}{{\left({16\pi^{2}}\right)^{2}}}\left\{ {162\left({\ln^{2}(1-\cos\theta)+\ln^{2}(1+\cos\theta)}\right)}\right.\\
 & + & \left.{\left({162\times2\ln\left({\frac{{2p^{2}}}{{m^{2}}}}\right)-108\ln\left({\frac{{2s_{0}^{3}}}{{m^{6}}}}\right)}\right)\ln(1-\cos^{2}\theta)+(\theta-\textup{independent terms})}\right\} \\
 & = & \frac{1}{{\left({16\pi^{2}}\right)^{2}}}\left\{ {162\left({\ln^{2}(1-\cos\theta)+\ln^{2}(1+\cos\theta)}\right)}\right.\\
 & + & \left.{\left({\frac{{162\times2}}{3}\ln\left({\frac{{8p^{6}}}{{m^{6}}}}\right)-108\ln\left({\frac{{2s_{0}^{3}}}{{m^{6}}}}\right)}\right)\ln(1-\cos^{2}\theta)+(\theta-\textup{independent terms})}\right\} \\
 & = & \frac{1}{{\left({16\pi^{2}}\right)^{2}}}\left\{ {162\left({\ln^{2}(1-\cos\theta)+\ln^{2}(1+\cos\theta)}\right)}\right.\\
 & + & \left.{108\ln\left({\frac{{4p^{6}}}{{s_{0}^{3}}}}\right)\ln(1-\cos^{2}\theta)+(\theta-\textup{independent terms})}\right\} \end{eqnarray*}
Thus, \begin{eqnarray*}
A(s,t,u) & = & -6\tilde{g}-\frac{{\tilde{g}^{2}}}{{16\pi^{2}}}\left[{-18\left\{ {\ln\left({\frac{{2s_{0}}}{s}}\right)+\ln\left({\frac{{s_{0}}}{t}}\right)+\ln\left({\frac{{s_{0}}}{u}}\right)}\right\} }\right]\\
 & - & \frac{{\tilde{g}^{3}}}{{\left({16\pi^{2}}\right)^{2}}}\left\{ {162\left({\ln^{2}(1-\cos\theta)+\ln^{2}(1+\cos\theta)}\right)}\right.\\
 & + & \left.{108\ln\left({\frac{{s^{3}}}{{16s_{0}^{3}}}}\right)\ln(1-\cos^{2}\theta)+(\theta-\textup{independent terms})}\right\} +o(\tilde{g}^{4})\end{eqnarray*}
Suppose, we choose the renormalization scale $s_{0}=
0.1\Lambda^{2}$. The angular dependence of the relevant matrix
element of $S_{3}$ is determined by\begin{eqnarray*}
g\left(\theta\right) & = &
\left[2\left[\ln(\frac{s}{^{_{0.2\times\sqrt[3]{2}\Lambda^{2}}}})\right]\ln(1-\cos^{2}\theta)+\ln^{2}(1+\cos\theta)+\ln^{2}(1-\cos\theta)\right]\end{eqnarray*}
we define,\begin{eqnarray*} a'_{2} & = & \frac{5}{2}\int\limits
_{-1}^{1}{d\cos\theta}P_{2}(\cos\theta)g\left(\theta\right)\end{eqnarray*}
We find,\begin{eqnarray*}
a'_{2} & = & \frac{5}{2}\times\frac{49}{18}+2\ln(\frac{s}{0.252\Lambda^{2}})\left(\frac{-5}{3}\right)\\
 & = & 6.81-3.33ln(15.87\frac{p^{2}}{\Lambda^{2}})\end{eqnarray*}
\begin{table}[htb]
putting in some values for $p^{2}/\Lambda^{2}$, we find
\begin{center}
\begin{tabular}{|c|c|c|c|c|}
\hline $\frac{p^{2}}{\Lambda^{2}}$ & 0.1 & 0.2 & 0.4 &
0.8\tabularnewline \hline \hline $a'_{2}$ & 5.27 & 2.96 & 0.66 &
-1.65\tabularnewline \hline
\end{tabular}
\caption{Legendre coefficient $a_{2}'$ for some values of
$\frac{p^{2}}{\Lambda^{2}}$.}
    \label{T:dimens}
\end{center}
Contribution from $S_{1}S_{3}$ term in terms of Legendre
coefficient turns out to be
    \end{table}
 \[
a_{2}^{(4,1)}=\frac{81\tilde{g^{4}}}{64\pi^{4}}a_{2}'\] To compare
this particular $O\left(g^{4}\right)$ contribution to the
non-local term, we consider,\begin{eqnarray*}
\frac{a_{2}^{\left(4,1\right)}}{a_{2}^{nonlcal}} & = &
\frac{27\widetilde{g^{2}}}{16\pi^{2}}\frac{a'_{2}}{\left({\frac{{p^{2}}}{{\Lambda^{2}}}}\right)^{2}}\end{eqnarray*}
\begin{table}[htb]
with, $\frac{{6g}}{{16\pi^{2}}}=0.001$(comparable to
$\frac{\alpha}{4\pi}$ in electrodynamics), we tabulate the ratio
for different values of
$\frac{p^{2}}{\Lambda^{2}}=\frac{s}{4\Lambda^{2}}$:
\begin{center}
\begin{tabular}{|c|c|c|c|c|}
\hline $\frac{p^{2}}{\Lambda^{2}}$ & 0.1 & 0.2 & 0.4 &
0.8\tabularnewline \hline \hline
$\frac{a_{2}^{\left(4,1\right)}}{a_{2}^{nonlcal}}$ & 0.062 &
0.0088 & 0.0005 & -0.0003\tabularnewline \hline
\end{tabular}
\caption{Comparison of $S_{1}S_{3}$-type terms with the leading
non-local contribution.}
    \label{T:dimens}
    \end{center}
     We saw earlier in section 3 that it was
possible to discern CV for $\frac{p^{2}}{\Lambda2}\gtrsim 0.2$. In
the same range of momenta, we find that contribution of this
$O\left(\widetilde{g}^{4}\right)$ term small enough to be ignored.
    \end{table}
\subsection{\textsc{the}$\left|{S_{2}}\right|^{2}$ \textsc{term}}
The contribution of this term is, \begin{eqnarray*}
 & = & \left({\frac{{9\widetilde{g}^{2}}}{{8\pi^{2}}}\ln\frac{{stu}}{{2s_{0}^{3}}}}\right)^{2}\\
 & = & \left({\frac{{9\widetilde{g}^{2}}}{{8\pi^{2}}}}\right)^{2}\left[{\ln\frac{{8p^{6}}}{{s_{0}^{3}}}+\ln(1-\cos^{2}\theta)}\right]^{2}\\
 & = & \left({\frac{{9\widetilde{g}^{2}}}{{8\pi^{2}}}}\right)^{2}\left[{\ln^{2}\frac{{s^{3}}}{{8s_{0}^{3}}}+\ln^{2}(1-\cos^{2}\theta)+2\ln\frac{{s^{3}}}{{8s_{0}^{3}}}\ln(1-\cos^{2}\theta)}\right]\\
 & = & \left({\frac{{9\widetilde{g}^{2}}}{{8\pi^{2}}}}\right)^{2}\left[{\ln^{2}\left({\frac{{s^{3}}}{{8s_{0}^{3}}}}\right)+\ln^{2}(1-\cos\theta)+\ln^{2}(1+\cos\theta)+2\ln(1-\cos\theta)\ln(1+\cos\theta)+}\right.\\
 &  & \left.{2\times3\ln\left({\frac{s}{{2s_{0}}}}\right)\ln(1-\cos^{2}\theta)}\right]\end{eqnarray*}
 The relevant angular dependent part is given below: \begin{eqnarray*}
f(\theta) & = & \left[{\ln^{2}(1-\cos\theta)+\ln^{2}(1+\cos\theta)+2\ln(1-\cos\theta)\ln(1+\cos\theta)+}\right.\\
 &  & \left.{2\times3\ln\left({\frac{s}{{2s_{0}}}}\right)\ln(1-\cos^{2}\theta)}\right]\end{eqnarray*}
Defining \[ a''_{2}=\frac{5}{2}\int\limits _{-1}^{+1}{d\cos\theta
P_{2}(\cos\theta)f(\theta)},\]
 we obtain, \begin{eqnarray*}
a''_{2} & = & \frac{5}{2}\times\frac{{49}}{{18}}-\frac{{25}}{{18}}+2\left({3\ln\frac{{s}}{{0.2\Lambda^{2}}}}\right)\left({\frac{{-5}}{3}}\right)\\
 & = & 5.42-10\left({\ln\frac{20p^{2}}{{\Lambda^{2}}}}\right)\end{eqnarray*}
\begin{table}[htb]
 We complete the table of $\frac{p^{2}}{{\Lambda^{2}}}$ versus $a''_{2}$.
\begin{center}
\begin{tabular}{|c|c|c|c|c|}
\hline $\frac{p^{2}}{{\Lambda^{2}}}$ & $0.1$ & $0.2$ & $0.4$ &
$0.8$\tabularnewline \hline $a''_{2}$ & -1.51 & -8.44 & -15.4 &
-22.3\tabularnewline \hline
\end{tabular}
\par
\caption{Legendre coefficient $a_{2}''$ for some values of
$\frac{p^{2}}{\Lambda^{2}}$ .}
    \label{T:dimens}
    \end{center}
$\left|{S_{2}}\right|^{2}$ term contributes the following Legendre
coefficient:
\end{table}
\begin{eqnarray*} a_{2}^{\left(4,2\right)} & = &
\frac{{81\tilde{g}^{4}}}{{64\pi^{4}}}\times a''_{2}\end{eqnarray*}
 Now, adding the Legendre coefficients to get the total contribution
to $\mathcal{R}$ in $O\left(g^{4}\right)$: \[
a_{2}^{\left(4\right)}=a_{2}^{\left(4,1\right)}+a_{2}^{\left(4,2\right)}=\frac{{81\tilde{g}^{4}}}{{64\pi^{4}}}\left({a'_{2}+a''_2}\right)\]
 Comparison of nonlocal effects of $O\left(g^{2}\right)$ and local
terms of next order is facilitated by looking at the ratio $r$: \[
r=\frac{{a_{2}^{\left(4\right)}}}{{a_{2}^{nonlocal}}}=\frac{{27\tilde{g}^{2}}}{{16\pi^{2}}}\frac{{\left({a'_{2}+a_{2}''}\right)}}{{\left({\frac{{p^{2}}}{{\Lambda^{2}}}}\right)^{2}}}\]
\begin{table}[htb]
We tabulate $r$ for various $\frac{p^{2}}{\Lambda^{2}}$ and with
$\frac{{6\tilde{g}}}{{16\pi^{2}}}=10^{-3}$:
\begin{center}
\begin{tabular}{|c|c|c|c|c|}
\hline $\frac{p^{2}}{\Lambda^{2}}$ & 0.1 & 0.2 & 0.4 &
0.8\tabularnewline \hline \hline $a_{2}^{\left(4\right)}$ &
2.34$\times 10^{-8}$ & -3.42$\times 10^{-8} $& -9.19$\times
10^{-8}$ & -1.49$\times 10^{-7}$\tabularnewline \hline \hline
$|r|$ & 0.04 & .02 & 0.01 & 0.004\tabularnewline \hline
\end{tabular}
\caption{Final comparison of neglected terms of
$O\left(\widetilde{g}^{4}\right)$ in (\ref{eq:CVO}) with the CV
amplitude.} \label{T:dimens}
\end{center}
 Thus, the contribution from the
terms of $O\left(g^{4}\right)$ we neglected is indeed a few
percent at best in this range of $\frac{p^{2}}{\Lambda^{2}}$ and
couplings.
\end{table}
\section{\textsc{\normalsize conclusions}}
We argued that physical theories may develop a small causality
violation at high enough energies; which could be due to diverse
causes such as a fundamental length scale, composite structure of
standard model particles etc. We wanted to study how it can be
observed experimentally. We considered as a model theory, the
nonlocal scalar theory, which embodies quantum violations of
causality. We demonstrated that CV could be observed by usual
laboratory measurements which obtain $\frac{d\sigma}{d\Omega}$ for
the exclusive elastic process $\phi\phi\rightarrow\phi\phi$.
Analysis of local contribution versus the non-local CV amplitude
enabled one to conclude that CV effects can be noticeable at
$s\sim\Lambda^{2}$ where $\Lambda$ is the large mass scale present
in the theory and a way to demonstrate its existence is via an
analysis of the angular distribution of scattering cross-section.
We constructed an observable that would serve the purpose if
higher order effects are negligible. We analyzed these
$O\left(g^{4}\right)$ terms and demonstrated that they are indeed
negligible compared to the CV terms at energies $s\leq\Lambda^{2}$
and for a typical coupling comparable to electromagnetic coupling
$\frac{\alpha}{4\pi}$. A work, along the same lines, but
applicable to the realistic cases of experimentally observed
exclusive processes $e^{+}e^{-}\rightarrow e^{+}e^{-}$,
$e^{+}e^{-}\rightarrow\mu^{+}\mu^{-}$ and
$e^{+}e^{-}\rightarrow\tau^{+}\tau^{-}$ is in progress.
\\
\\
{\textbf{ACKNOWLEDGEMENT}}\\
AH would like to thank 	NISER, Bhubaneswar for support where a part of the work was done.\\
\\


\begin{thebibliography}{10}
\bibitem{PU} \textbf{A. Pais and G. E. Uhlenbeck, Phys. Rev. 79,
145-165 (1950). }
\bibitem{J06} \textbf{S. D. Joglekar, hepth/0601006}
\bibitem{NA}\textbf{ See, e.g. K. Namsrai,} \textbf{\emph{Nonlocal
Quantum Field Theory and Stochastic Quantum Mechanics}}
\textbf{(D. Reidel Publishing company). }
\bibitem{M90}\textbf{ J.Moffat Phys. Rev. D41,1177(1990) }
\bibitem{NC}\textbf{ See e.g. N. Seiberg, Leonard Susskind, N. Toumbas;
JHEP 0006:044,(2000) }
\bibitem{KW92}\textbf{ G. Kleppe and R. P. Woodard, Nucl. Phys. B388,
81 (1992) }
\bibitem{EW}\textbf{ D. A. Eliezer, and R. P. Woodard, Nucl.Phys.B
325, 389 (1989). }
\bibitem{E91}\textbf{ E. D. Evens et al, Phys Rev D43, 499 (1991) }
\bibitem{j01}\textbf{ S.D.Joglekar, J. Phys. A34, 2765-2776 (2001) }
\bibitem{j01_2}\textbf{ S.D.Joglekar,Int.J.Mod. Phys.A 16, (2001). }
\bibitem{js}\textbf{ S. D. Joglekar and G. Saini, Z. Phys. C.76,
343-353 (1997); A. Basu, and S. D. Joglekar, J. Math. Phys. 41,
7206-7219 (2000). }
\bibitem{CO92}\textbf{N. J. Cornish, Int.J.Mod. Phys.A 7, 6121-6157
(1992). }
\bibitem{JJ}\textbf{ A. Jain and S.D.Joglekar, Int.J. Mod. Phys.A
19, 3409(2004) }
\bibitem{J08} S.D. Joglekar, Int. J. of Theo. Phys.47, (2008)
\bibitem{BS}\textbf{N.N.Bogolibov and D.V.Shirkov, `Introduction
to Theory of Quantized Fields' ($3^{rd}$ ed.), John Wiley(1980).
See pg. 200-220. } \textbf{\bibitem{MDR}P.N. Maher, L. Durand, K.
Reiselmann Phys.Rev. D 48,1061(1993)} \textbf{\bibitem{PS} M. E.
Peskin, and D. V. Schroeder, An introduction to quantum field
theory (Westview, Boulder, Colo., 2003)}
\bibitem{AJ} A. Ayyer and S.D.Joglekar (Unpublished work)
\end{thebibliography}
\end{document}